\documentclass[pra,aps,twocolumn,superscriptaddress,showpacs,10pt,floatfix]{revtex4}

\usepackage{amsmath}
\usepackage{amssymb}
\usepackage{graphicx}
\usepackage{microtype}
\usepackage{times}
\usepackage[svgnames]{xcolor}
\usepackage[pdftex]{hyperref}
\hypersetup{colorlinks=true,linkcolor=blue,citecolor=ForestGreen,urlcolor=DarkOrchid}

\graphicspath{{./}}

\newcommand{\sectionname}{Section}

\newcommand{\kff}{$\text{K}_{4,4}$ }

\def\DWQ{\mbox{D-Wave 2000Q$\ $}}
\def\TTS{\text{TTS}}

\newcommand{\nasa}{\affiliation{Quantum Artificial Intelligence Lab., NASA Ames Research Center, Moffett Field, CA 94035, USA}}
\newcommand{\sgt}{\affiliation{Stinger and Ghaffarian Technologies, 7701 Greenbelt Road, Suite 400 Greenbelt, MD 20770, USA}}
\newcommand{\tamu}{\affiliation{Department of Physics and Astronomy, Texas A\&M University,  College Station, Texas 77843-4242, USA}}

\newcommand{\sfi}{\affiliation{Santa Fe Institute, 1399 Hyde Park Road, Santa Fe, New Mexico 87501 USA}}
\newcommand{\oqbit}{\affiliation{1QB Information Technologies (1QBit), Vancouver, British Columbia, Canada V6B 4W4}}
\newcommand{\googlenyc}{\affiliation{Google Inc., 111 8th Avenue New York, NY 10011, USA}}

\begin{document}

\title{The pitfalls of planar spin-glass benchmarks: Raising the bar for quantum annealers (again)}

\author{Salvatore Mandr{\`a}}
\email{salvatore.mandra@nasa.gov}
\nasa
\sgt

\author{Helmut G.~Katzgraber}
\email{hgk@tamu.edu}
\tamu
\oqbit
\sfi

\author{Creighton Thomas}
\email{creightonthomas@gmail.com}
\googlenyc

\date{\today}

\pacs{75.50.Lk, 75.40.Mg, 05.50.+q, 03.67.Lx}

\begin{abstract}

In an effort to overcome the limitations of random spin-glass benchmarks
for quantum annealers, focus has shifted to carefully-crafted
gadget-based problems whose logical structure has typically a planar
topology. Recent experiments on these gadget problems using a
commercially-available quantum annealer have demonstrated an impressive
performance over a selection of commonly-used classical optimization
heuristics. Here we show that efficient classical optimization
techniques, such as minimum-weight perfect matching, can solve these
gadget problems exactly and in polynomial time. We present approaches on
how to mitigate this shortcoming of commonly-used benchmark problems
based on planar logical topologies.

\end{abstract}

\maketitle

\section{Introduction}
\label{sec:intro}

The quest for quantum speedup using analog quantum annealing machines
with a transverse field remains elusive. There have been multiple
attempts
\cite{ronnow:14a,boixo:14,katzgraber:15,venturelli:15a,denchev:16} to
demonstrate that the D-Wave Systems Inc.~quantum annealers can
outperform classical optimization methods. Unfortunately, it has been
relatively straightforward for classical optimization algorithms to stay
ahead in this race \cite{mandra:16b}. Either the random spin-glass
benchmark problems were too easy on the quasi-planar topology of the
D-Wave quantum annealer \cite{ronnow:14a,katzgraber:14,katzgraber:15},
or the logical structure of carefully-crafted problems designed to give
the annealer an advantage have a trivial structure \cite{denchev:16}.

The notable advances made by D-Wave Systems Inc.~in the development of
medium-scale quantum annealing technologies has inspired multiple
corporations (e.g., Microsoft, Google, and IBM) to further invest in quantum
computing, in addition to large-scale government-funded
projects. The \DWQ device is a special-purpose quantum
optimization machine specialized in minimizing quadratic unconstrained
binary cost functions by means of quenching quantum fluctuations induced
by a transverse field. The generic cost function to be minimized is
given by
\begin{equation}
{\mathcal H}_{\rm P} = 
	\sum_{i = 1}^n J_{ij} \sigma_i \sigma_j + 
	\sum_{i = 1}^n h_i \sigma_i,
\label{eq:isg}
\end{equation}
where the $n$ variables $\sigma_i \in \{\pm 1\}$ are Boolean and the
couplings $J_{ij} \in {\mathbb R}$ and biases $h_i \in {\mathbb R}$ are
the parameters that define the problem to be minimized. In the case of
the D-Wave chip, these variables are arranged in the so-called Chimera
topology \cite{bunyk:14}. Real-world applications then require the
embedding of the problems onto this topology, thus typically resulting
in an embedding overhead that results in logical problems with less
sites than the native topology of the chip. As such, it is desirable to
find classes of problems for benchmarking that ideally use the complete
set of variables on the chip, while not being a trivial optimization
problem \cite{ronnow:14a}.

There have been multiple approaches to harden the benchmark problems to
be solved on different generations of the D-Wave device, ranging from
post-selection methods based on statistical-physics metrics
\cite{katzgraber:15} to planting of solutions \cite{hen:15a} and the
engineering of problems based on the classical algorithmic complexity of
the Hamze--de Freitas--Selby \cite{hamze:04,selby:14} algorithm
\cite{marshall:16}. Although these approaches generated harder problems
for different generations of the D-Wave devices and there were some
suggestions that there is some level of ``quantumness'' in the device
\cite{katzgraber:15}, only studies tailored to explicitly demonstrate
quantumness, as well as attempt to determine (quantum) speedup
\cite{ronnow:14a,mandra:16b} pushed the field forward noticeably. Both
the studies of Denchev {\em et al.}~\cite{denchev:16} and King {\em et
al.}\cite{king:17} focused on the generation of logical problems
designed to elucidate the value of quantum fluctuations, as well as to
``fool'' archetypal classical optimization algorithms (e.g., simulated
annealing \cite{kirkpatrick:83}, the classical pendant to quantum
annealing \cite{kadowaki:98,farhi:00}) to become stuck in the
carefully-designed energy landscape of the problems. In
Ref.~\cite{mandra:16b} it was subsequently shown that the use of
state-of-the-art optimization techniques beyond simulated annealing for
the benchmarks designed in Ref.~\cite{denchev:16} resulted in
time-to-solutions scaling considerably better than the D-Wave device, as
well as simulated quantum annealing. In this work we demonstrate that if
the logical problems to be optimized on the D-Wave device have a planar
structure, a quantum annealer would have to scale polynomially in the
number of (logical) variables (i.e., be exponentially faster) to compete
with the current classical state-of-the-art for frustrated problems on
planar topologies, such as the minimum-weight perfect-matching (MWPM)
exact algorithm \cite{hartmann:01,hartmann:04}. We emphasize that {\em
both} the benchmark instances designed by Denchev {\em et
al.}~\cite{denchev:16} and King {\em et al.}\cite{king:17} suffer from
the same problem. Namely, they can both be solved in polynomial time.
Although one could, in principle, compare the quantum annealer against
fast heuristics such as the Hamze--de Freitas--Selby
\cite{hamze:04,selby:14} algorithm \cite{marshall:16} or parallel
tempering Monte Carlo with isoenergetic cluster moves \cite{zhu:15b}, if
claims of speedups over many orders of magnitude against classical
algorithms are made, then the true state-of-the-art for planar
topologies should be included in the study.

Although one might argue that exploiting the logical structure of the
problem could be seen as ``cheating,'' combining MWPM techniques with
simple cluster-finding and/or decimation techniques that are also
polynomial in the size of the input would still scale exponentially
faster than the D-Wave device. However, there would be no more guarantee
for an {\em exact} result and the cluster-detecting MWPM algorithm
could, at best, be seen as a heuristic that scales polynomial in the
size of the input.

The paper is structured as follows. In \sectionname~\ref{sec:model} we
describe the crafted benchmark problem designed by D-Wave Systems
Inc.~\cite{king:17} and in \sectionname~\ref{sec:methods} we describe the
classical algorithms and methods we used in our analysis of these
problems.  Results are summarized in \sectionname~\ref{sec:results},
followed by a discussion that also includes different strategies to
design problems on quantum annealing machines that might have potential
for quantum speedup and cannot be solved with polynomial algorithms for
planar technologies.

\begin{figure*}[t!]
\includegraphics[width=0.90\columnwidth]{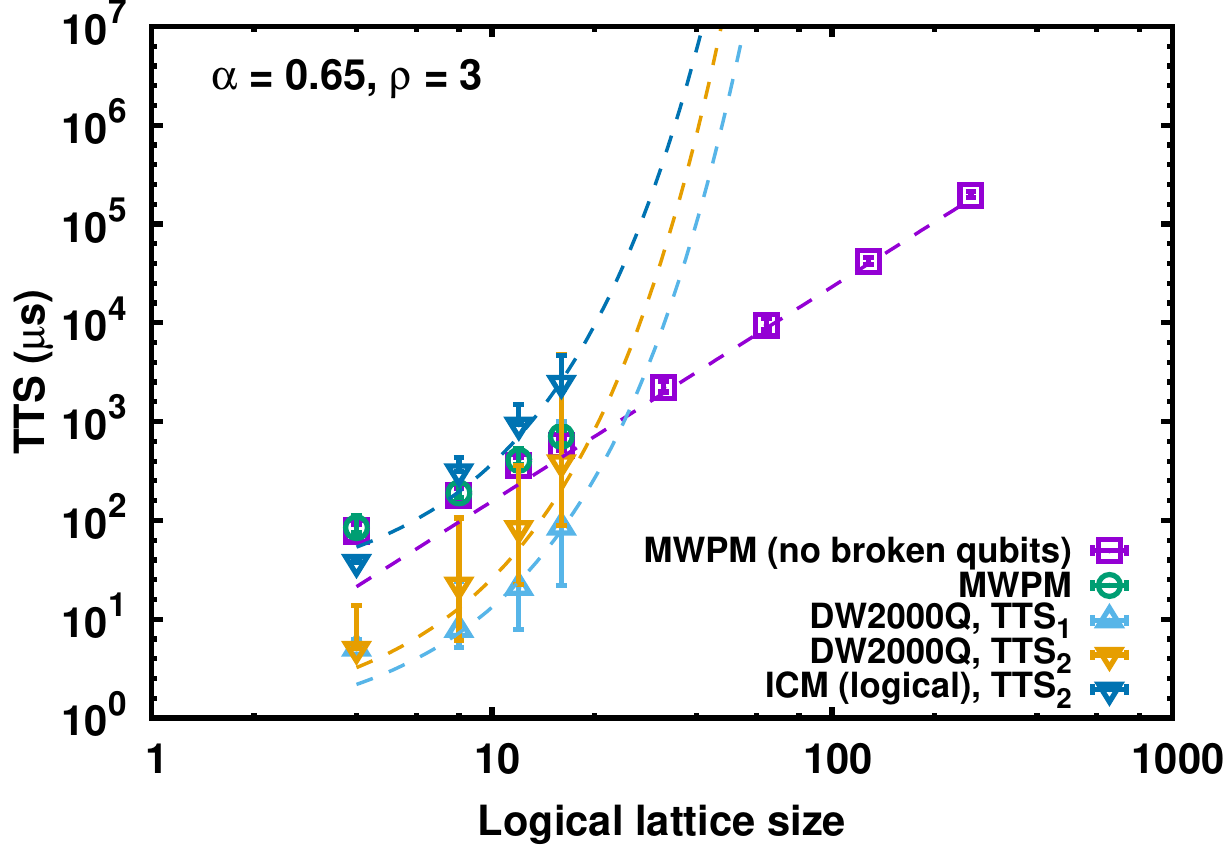}
\hspace*{1em}
\includegraphics[width=0.90\columnwidth]{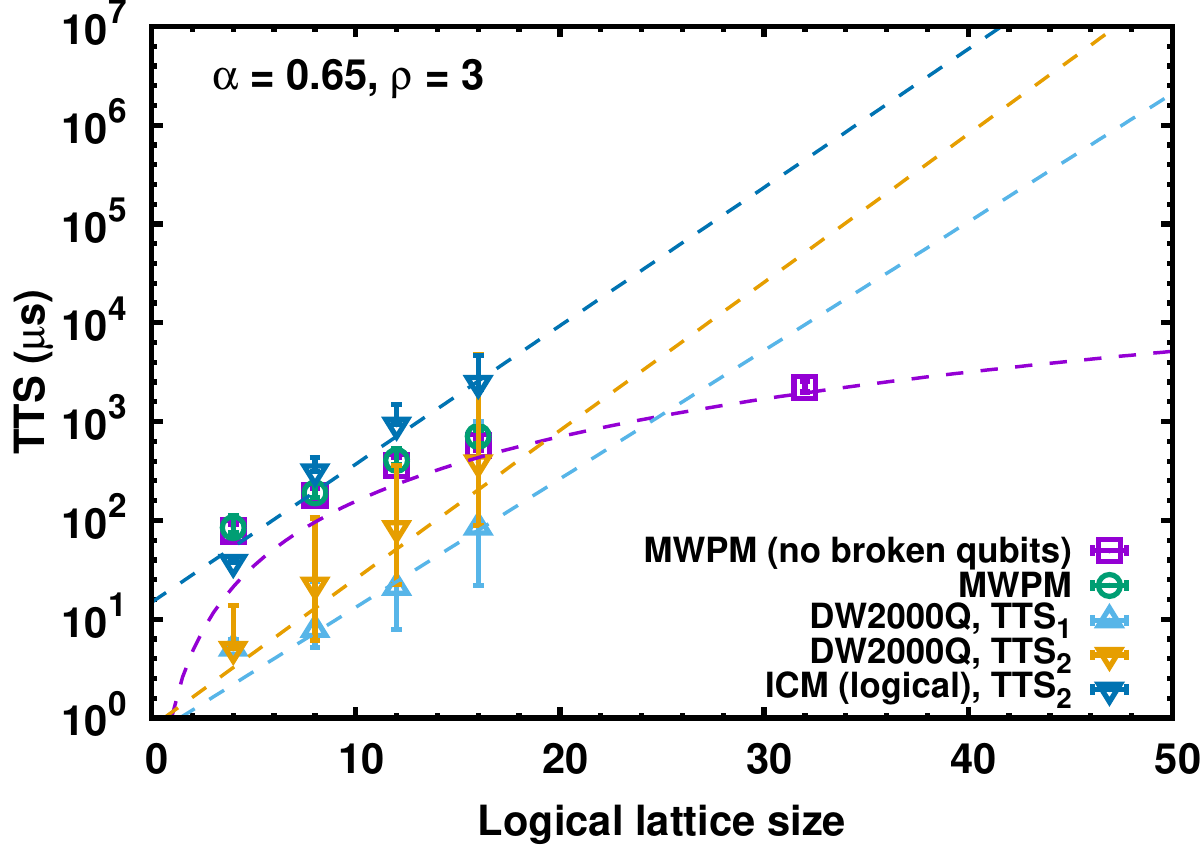}\\[0.5em]
\includegraphics[width=0.90\columnwidth]{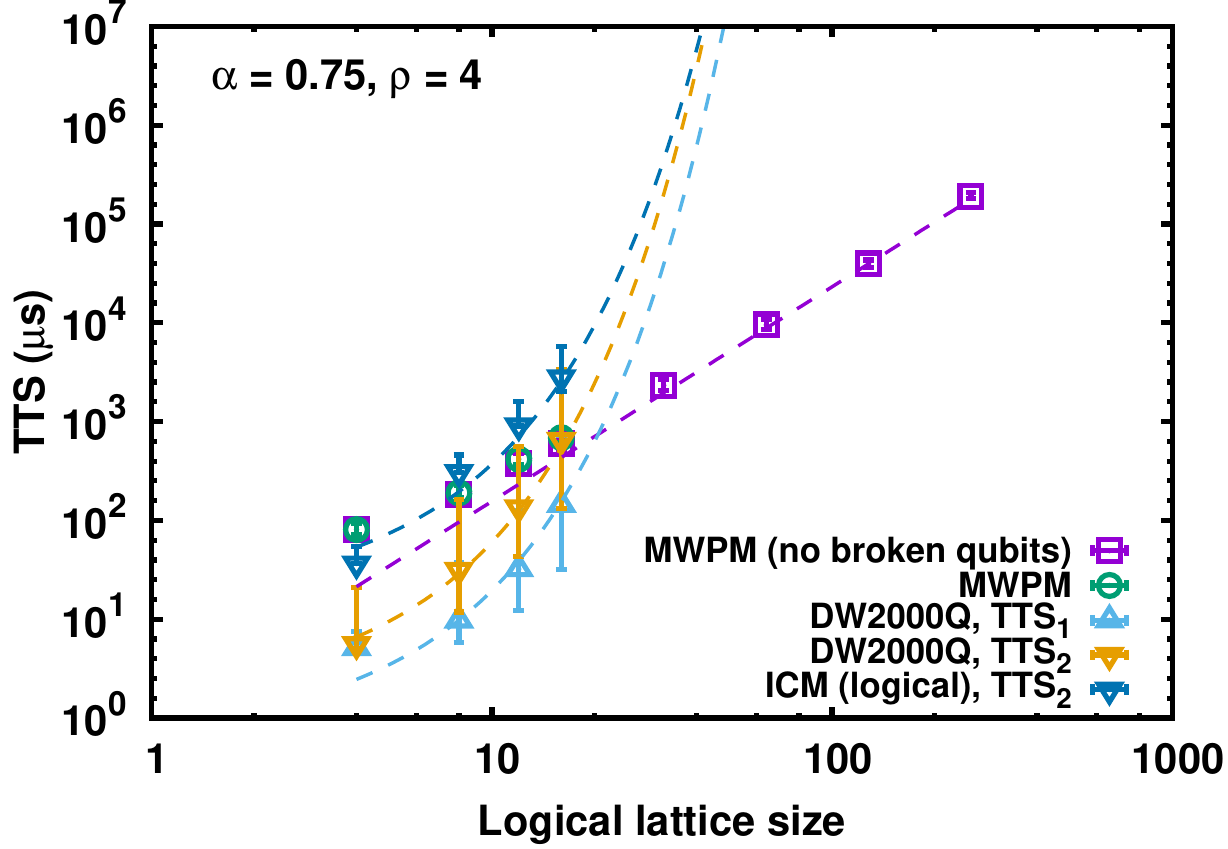}
\hspace*{1em}
\includegraphics[width=0.90\columnwidth]{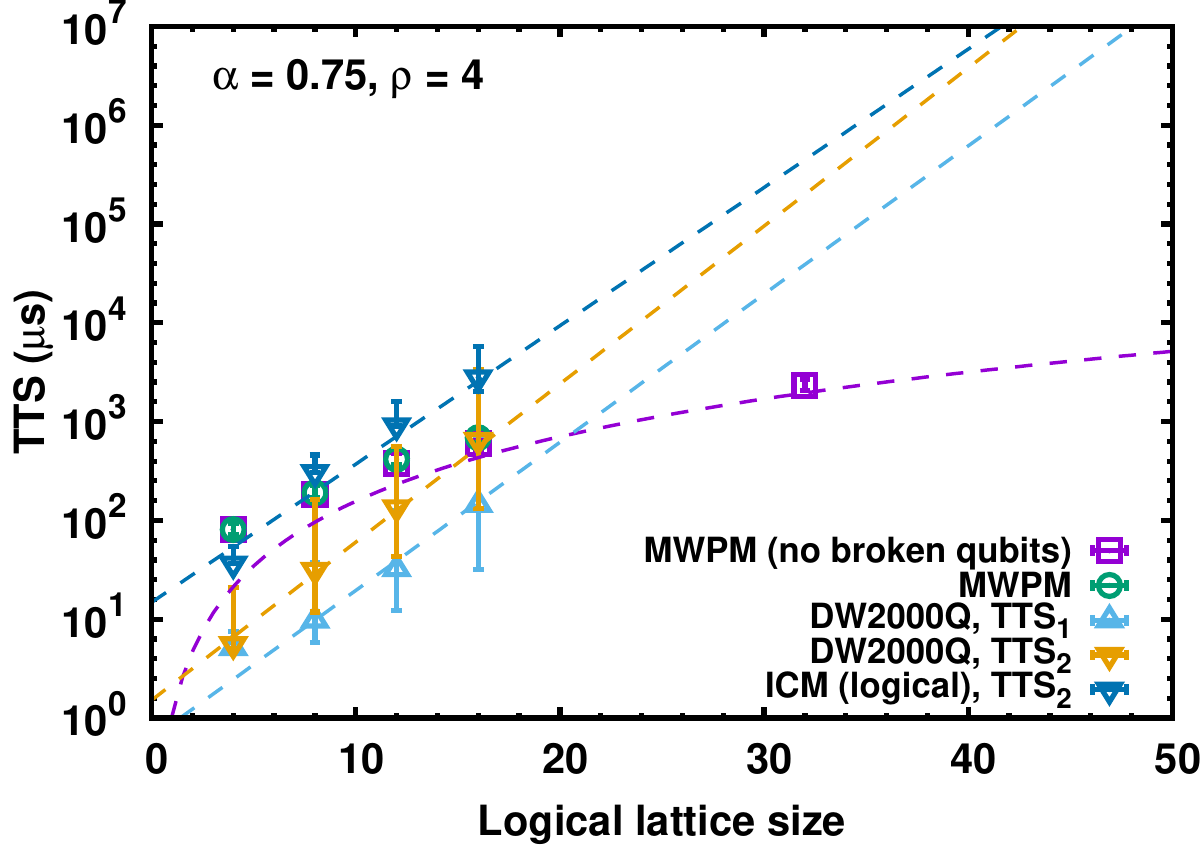}\\[0.5em]
\includegraphics[width=0.90\columnwidth]{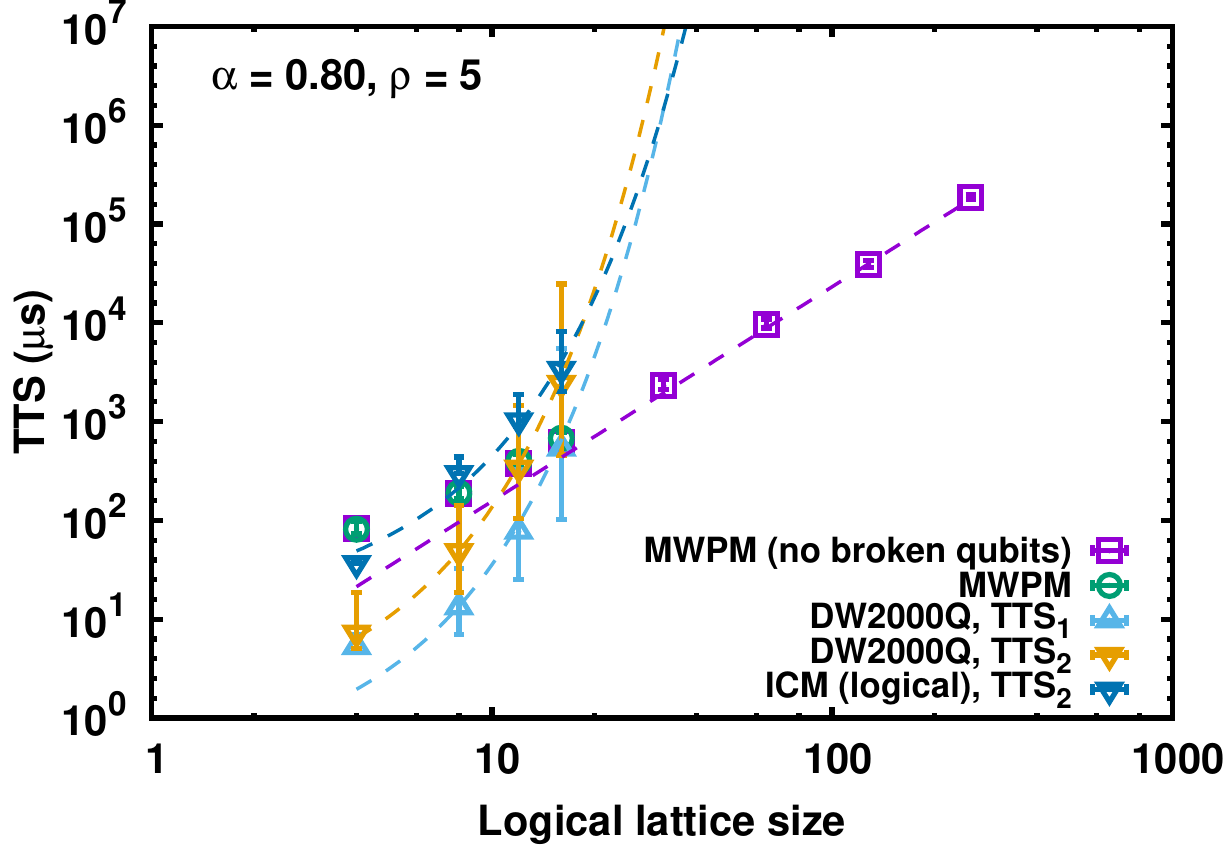}
\hspace*{1em}
\includegraphics[width=0.90\columnwidth]{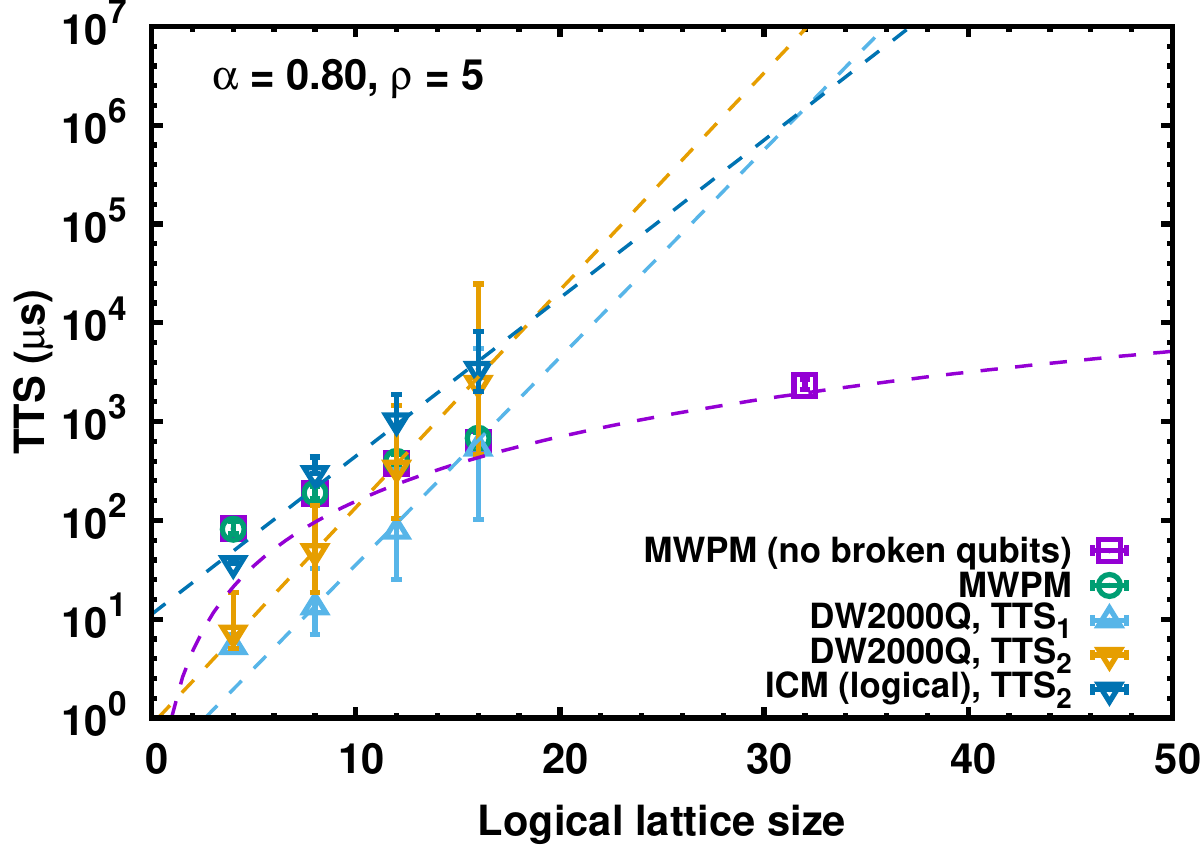}\\[0.5em]
\includegraphics[width=0.90\columnwidth]{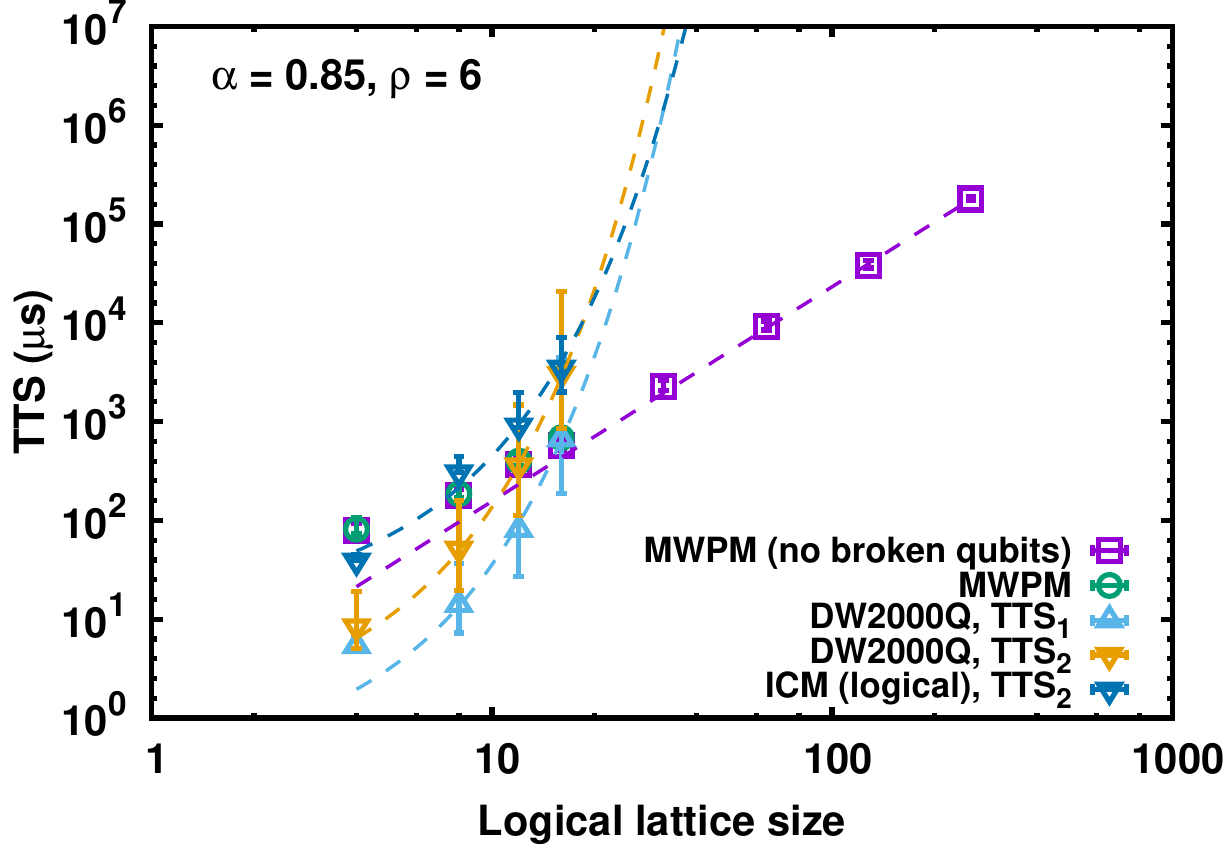}
\hspace*{1em}
\includegraphics[width=0.90\columnwidth]{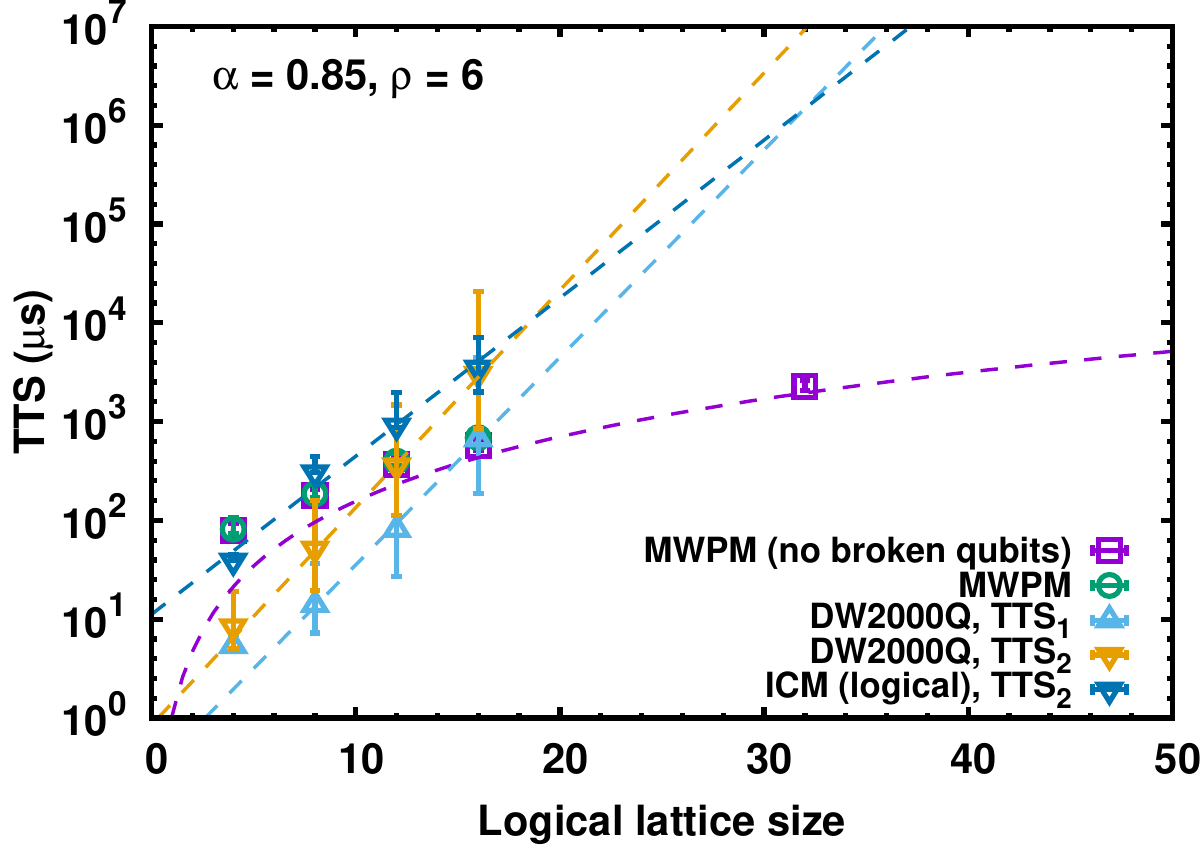}
\caption{\label{fig:scaling}
Scaling of the TTS in $\mu$s as a function of logical variables in a
log-log scale (left) or linear-log scale (right). Data for the \DWQ
(DW2000Q) quantum annealer for both definitions of the TTS are compared
to MWPM and ICM. Because the maximal logical problem size is limited on
the \DWQ to $16\times16$ variables, we have generated artificial full
Chimera lattices with no broken qubits of up to $256\times256$ \kff unit
cells. Note that MWPM scales linearly in a log-log scale, whereas the
\DWQ scales exponentially. In all panels, data points represent the
$50\%$ of the TTS, while error bars represent the $5\%$-$95\%$ of the
distribution.  Although the \DWQ is relatively fast for a small number
of logical qubits $n$, MWPM quickly outperforms the quantum device by
several orders of magnitude for the larger lattice sizes. Raw \DWQ data
taken from Ref.~\cite{king:17}.}
\end{figure*}

\section{D-Wave's crafted problems}
\label{sec:model}

Given its hardware limitations, not all possible couplings $\{J_{ij}\}$
between two arbitrary qubits $i$ and $j$ can be set in the D-Wave
quantum annealer. Indeed, only those couplings belonging to the native
Chimera topology can be independently tuned within the range
$[-1,\,+1]$, while the remaining are set to zero. The Chimera topology
\cite{bunyk:14} is composed of $k \times k$ unit cells, each containing
a \kff fully-connected bipartite graph of $8$ qubits. The unit cells are
coupled together so that only adjacent unit cells share couplings.
Despite the somewhat restrictive structure of the lattice, it can be
shown that, in principle, any topology can be embedded, albeit at the
cost of using multiple physical variables to define a logical variable.

In Ref.~\cite{king:17}, the latest incarnation of the quantum annealer,
namely the \DWQ with over $2000$ quantum bits, is tested using a set of
carefully crafted optimization problems also referred to as the
``frustrated cluster loop'' (FCL) problems.  One of the main
characteristics which makes the FCL problems appealing for benchmarking
is that many classical heuristics struggle with minimizing the value of
the cost function, even though the optimal configuration can be deduced
by exploiting the actual structure of the problem
\cite{hen:15a,king:17}.

Although the FCL problems can, in principle, be directly generated for
the Chimera topology \cite{hen:15a}, the D-Wave Systems Inc.~group has
chosen a slightly different strategy, divided into two steps, which
assists in elucidating the effects of the landscape ruggedness:
\begin{enumerate}

\item All couplings inside a \kff unit cell are set to be ferromagnetic,
i.e., $J_{ij} = -1$, $\forall i,j \in$ \kff. because the unit cells are
fully connected, all the ``physical'' qubits within a single cell are
forced to behave as a single ``logical'' qubit. This process generates a
two-dimensional lattice with open boundary conditions of these logical
variables.

\item The FCL instances are then generated on the {\em logical} topology
with a varying level of ruggedness of the energy landscape and
parameters $\alpha$ (clause-to-variable ratio) and $\rho$ (precision),
as defined in Ref.~\cite{king:15}. Note that for the ruggedness $R$ we
expect $R \ge \rho$.

\end{enumerate}
This local ruggedness $R$ then makes the problems hard to treat for
typical classical optimization techniques with the interactions for the
logical qubits being in the range $[-R,\,+R]$. Disconnected graphs are
discarded in the study.

It is important to stress that, despite the fact that \DWQ can optimize
problems on the Chimera topology, the benchmark problems used in
Ref.~\cite{king:17} are defined on the \emph{logical topology} of the
machine, namely on a two-dimensional lattice with open boundaries.
Therefore, a fair comparison requires that the \DWQ benchmarking be
performed against heuristics which are optimized for the \emph{logical
problem}, rather than on the Chimera topology. It is noteworthy,
however, that the \DWQ solves problems on the full Chimera lattice,
i.e., the machine seems to be able to efficiently overcome local energy
barriers.

\section{Methods}
\label{sec:methods}

In this \sectionname, we briefly outline the algorithms used, as well as
the definition of time-to-solution used in the benchmarks. Reference
\cite{zhu:15b} provides the necessary details for the isoenergetic
cluster move (ICM) heuristic.

\subsection{Minimum-weight perfect matching algorithm}
\label{ssec:mwpm}

The minimum-weight perfect matching (MWPM) algorithm is an exact
classical algorithm designed to find optimal configurations for planar
two-dimensional spin-glass-like optimization problems without biases
(i.e., $h_i = 0$, $\forall i \in n$). The algorithm is polynomial in the
size of the input $n$.  The MWPM algorithm consists of three steps:
\begin{enumerate}

\item The planar spin-glass problem is mapped onto a minimum-weight
perfect matching problem.

\item The minimum weight-perfect matching problem is solved exactly
using the deterministic Blossom algorithm \cite{kolmogorov:09} that
scales polynomially in the size of the input.

\item The perfect matching solution is then translated to the optimal 
configuration for the spin-glass problem.

\end{enumerate}

\begin{figure}[t!]
\includegraphics[width=0.45\columnwidth]{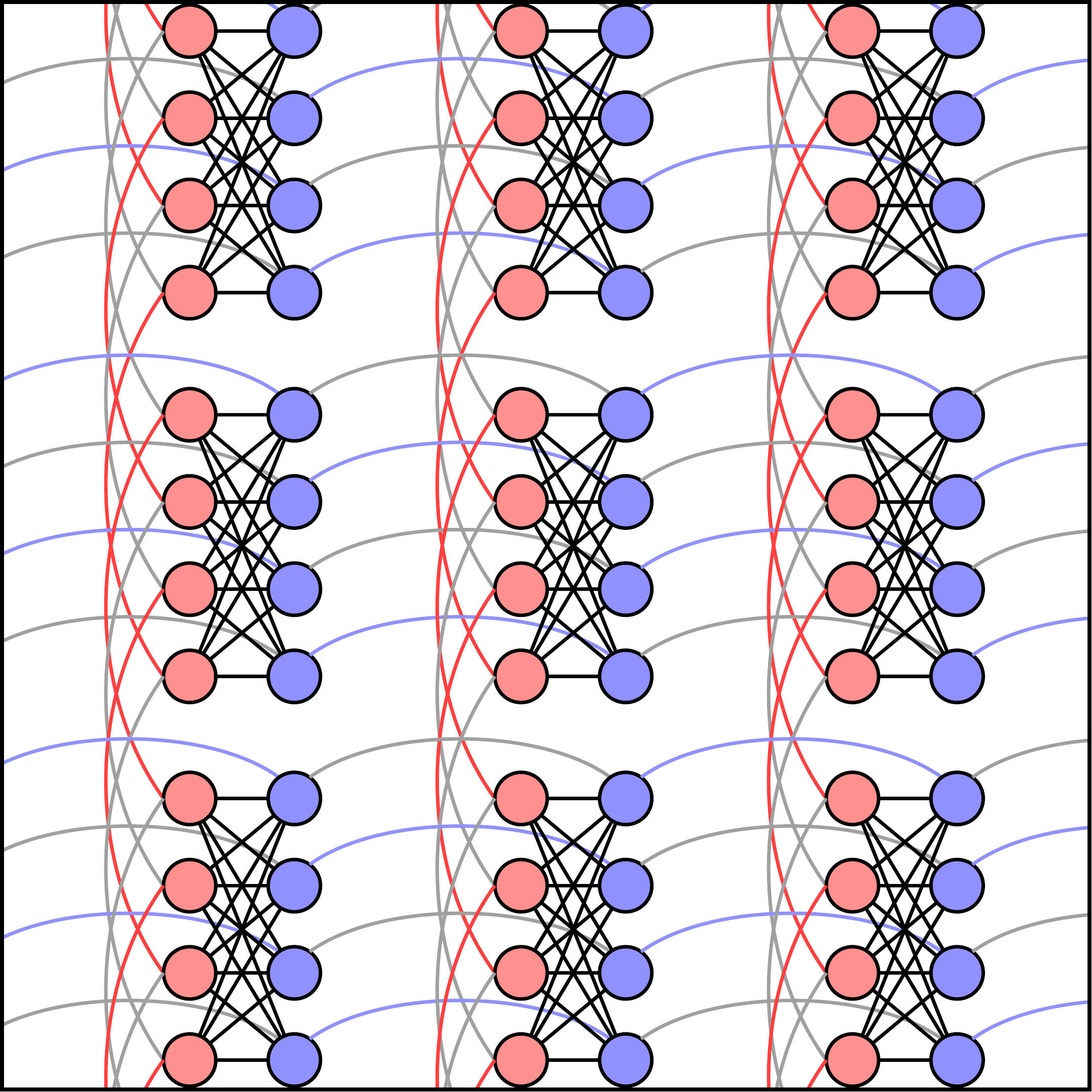}
\hspace*{1em}
\includegraphics[width=0.45\columnwidth]{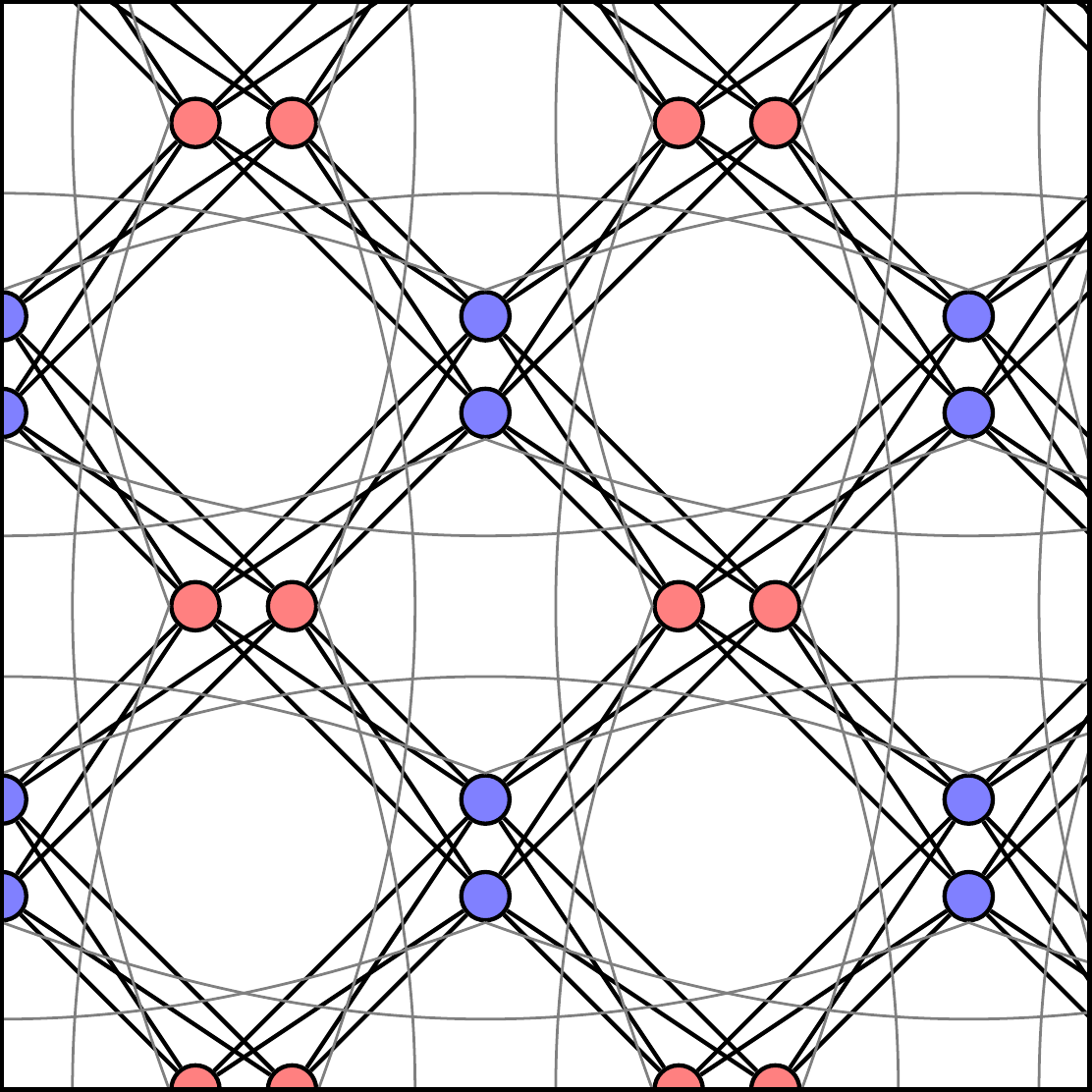}
\caption{\label{fig:ac}
Left panel: D-Wave Systems Inc.~Chimera topology \cite{bunyk:14}. To
generate the anticluster lattice (right panel), two qubits in different
\kff cells are contracted to a logical qubit with a strong ferromagnetic
coupler. Different shadings are used as a guide to the eye. Right panel:
Anticluster lattice of logical qubits. Each bulk logical qubit has $10$
neighbors and the lattice is nonplanar. Therefore, polynomial exact algorithms
cannot be used to solve the logical problems.}
\end{figure}

\subsection{Definition of time-to-solution}
\label{ssec:tts}

Heuristic methods, such as simulated annealing, simulated quantum
annealing, the \DWQ quantum annealer or isoenergetic cluster moves, can
only determine the optimum of the cost function up to a probability $p$.
If the optimization procedure requires a certain time $T$, given that
the optimum is only obtained with a probability $p$, it is necessary to
define a time-to-solution (TTS).  A simple (yet naive) possibility
consists of making the observation that, on average, one needs to repeat
the computation $\approx 1/p$ times in order to observe one optimal
result. Therefore, for a computational time $T$, a possible definition
of the TTS is
\begin{equation}\label{eq:tts1}
  \TTS_1 = \frac{T}{p}.
\end{equation}
A commonly-used, more accurate definition of the TTS that incorporates
the cost of having uncertainty in the heuristic results is given as
follows: Let $k$ be the number of (unknown) iterations required to have
a probability of success of at least $99\%$, i.e., $s = 0.99$. The
probability that all $k$ attempts fail to find the correct answer is
$P_\text{wrong} = (1 - p)^k$. Because an overall probability of success
$s$ is needed, it is required that $P_\text{wrong} < s$. Therefore, $k$
must be at least $k > \log(1 - s)/\log(1 - p)$. Assuming that each
attempt require $T$ times, the TTS can be defined as
\begin{equation}\label{eq:tts2}
  \TTS_2 = T\, \frac{\log(0.01)}{\log(1-p)}.
\end{equation}
Note that $\TTS_2 < \TTS_1$ at fixed $p$ and $T$. However, in general,
$\TTS_2$ is preferred, because it gives a lower bound to the overall
probability. Reference \cite{king:17} used the definition shown in
Eq.~\eqref{eq:tts1}. Using the raw data from Ref.~\cite{king:17}, we
have converted their results into the more commonly used TTS shown in
Eq.~\eqref{eq:tts2}.

\section{Results}
\label{sec:results}

Figure \ref{fig:scaling} summarizes our results where we compare the
performance of the \DWQ quantum annealer to both ICM and MWPM.  The
simulations were performed on a single core of an Intel(R) Xeon(R) CPU
(E5-1650v2 with 3.50GHz clock speed).  While both the \DWQ quantum
annealer and ICM scale exponentially, MWPM scales polynomially with the
size of the input. To show this in more detail, we have generated
artificial problems on perfect Chimera lattices of up to $256\times256$
logical variables. Although for small systems the \DWQ chip is
remarkably fast, only by doubling the largest number of logical
variables on the chip results in MWPM outperforming the quantum annealer
by approximately three orders of magnitude. One has to remember,
however, that the \DWQ quantum annealer is a special purpose machine
designed to minimize binary problems, whereas classical CMOS
technologies require other processes to run, such as the operating
system, kernel and other concurrent processes.

\section{Discussion}
\label{sec:discussion}

Although the D-Wave chip shows remarkable promise, in this work we show
that benchmarks which encode the logical problem on a planar topology is
bound to fail in reaching the crown of quantum speedup. The quantum
annealer would have to perform exponentially faster, in order to
outperform the exact polynomial algorithm.

So how can we eventually prove the value of quantum annealing
topologies? First and foremost, encode the problems in nonplanar logical
topologies to ensure that no exact polynomial algorithms can be used.
One possible approach we call ``anticlusters'' (see Fig.~\ref{fig:ac})
is to contract the links {\em between} the \kff cells to become logical
variables. This would results in a nontrivial nonplanar topology where
each logical variable has $10$ neighbors, except for the logical
variables on the edges of the lattice which only have $5$ neighbors. For
a Chimera lattice with $c \times c$ \kff cells (i.e., $8c^2$ physical
qubits), the corresponding anticluster lattice would have $4c(c+1)$
logical qubits arranged on a square-lattice-like structure of linear
dimensions $c \times c$. The large connectivity of the anticluster
lattice, as well as the large number of logical variables allows for the
generation of nontrivial benchmarking problems. For example, overlaying
this topology that resembles the offspring of a farm fence with a square
lattice with frustrated cluster loops or post-selected spin-glass
problems, should generate hard benchmarks for classical algorithms.

A second alternative to demonstrate the capabilities of the \DWQ is to
use the machine as a physical simulator to study nontrivial quantum
physics Hamiltonians \cite{harris:17x}. Because these are very hard to
simulate already for small numbers of variables, the \DWQ might be able
to outperform classical simulation techniques in the near future.

\begin{acknowledgments}

We would like to thank the D-Wave team and, in particular, J.~King for
sharing their problem instances and raw data, as well as comments on the
manuscript. We are also indebted to Richard Harris at D-Wave for sharing
potentially groundbreaking, yet hopefully soon-to-be-published results.
We thank H.~Munoz-Bauza for rendering the anticluster problems.
H.~G.~K.~would like to thank Zheng Zhu for sharing his implementation of
the ICM heuristic, as well as F.~Hamze and G.~Rosenberg for multiple
discussions. He also thanks Franziskaner Naturtr\"ub for inspiration.
H.~G.~K.~acknowledges support from the NSF (Grant No.~DMR-1151387).
S.~M.~acknowledges Eleanor~G.~Rieffel for the useful discussions and the
careful reading of the manuscript. This research is based upon work
supported in part by the Office of the Director of National Intelligence
(ODNI), Intelligence Advanced Research Projects Activity (IARPA), via
MIT Lincoln Laboratory Air Force Contract No.~FA8721-05-C-0002.  The
views and conclusions contained herein are those of the authors and
should not be interpreted as necessarily representing the official
policies or endorsements, either expressed or implied, of ODNI, IARPA,
or the U.S.~Government.  The U.S.~Government is authorized to reproduce
and distribute reprints for Governmental purpose notwithstanding any
copyright annotation thereon.

\end{acknowledgments}

\bibliography{refs.bib}

\end{document}